\documentclass[
 reprint,
 amsmath,amssymb,
 aps,
 pra,
]{revtex4-2}

\usepackage{amsmath,amssymb,amsfonts}
\usepackage{bm}
\usepackage{mathtools}
\usepackage{braket}
\usepackage{graphicx}
\usepackage[colorlinks=true,linkcolor=blue,citecolor=blue,urlcolor=blue]{hyperref}
\usepackage{comment}
\usepackage{tikz}
\usetikzlibrary{arrows.meta,calc,decorations.pathmorphing}
\usepackage{mathrsfs}
\usetikzlibrary{decorations.pathreplacing}

\def\dj{d\kern-0.4em\char"16\kern-0.1em}
\def\Dj{\mbox{\raise0.3ex\hbox{-}\kern-0.4em D}}

\newcommand{\ii}{\mathrm i}

\newcommand{\ketvac}{\lvert 0\rangle}

\begin{document}

\title{Bell nonlocality from twisted statistics}

\author{Ivana \Dj or\dj evi\'{c}}
\author{Jovan Potrebi\'{c}}
\author{Aleksandra Go\v{c}anin}
\author{Dragoljub Go\v{c}anin}
\email{dragoljub.gocanin@ff.bg.ac.rs}
\affiliation{Faculty of Physics, University of Belgrade, Studentski Trg 12--16, 11000 Belgrade, Serbia}

\begin{abstract}
We investigate Bell correlations for a free real quantum scalar field on the noncommutative Moyal plane. Although the free field dynamics and the one-particle sector remain unchanged, the deformation enters through twisted multiparticle statistics and its Fock-space dressing representation. A classical external source coupled locally to the twist-dressed quantum field prepares coherent superpositions of momentum-pair configurations propagating toward two spacelike-separated laboratories. The momentum-dependent twist phases are generally nonfactorizable and generate entanglement between the corresponding wave-packet modes. We show that suitable local mode measurements lead to a violation of the CHSH Bell inequality. The resulting correlations provide an operational probe of the noncommutative structure encoded in the multiparticle sector of the quantum field.

\end{abstract}

\maketitle

\section{Introduction}
\label{sec:introduction}

Bell's theorem places fundamental constraints on the correlations that can arise in a locally realistic physical description
\cite{Bell1964, Bell1976}. In a standard Bell scenario, two spacelike-separated
observers, Alice and Bob, independently choose measurement settings \(a\) and
\(b\), and obtain outcomes \(\alpha\) and \(\beta\), respectively. Their
systems may share a common preparation, or may have interacted in their joint
past, but the measurement events are arranged so that no causal signal can
propagate from one measurement region to the other during the experiment.
A locally realistic description introduces the so-called hidden variables \(\lambda\), distributed
according to \(q(\lambda)\), which characterize the common preparation. The
joint probabilities then admit the decomposition
\begin{equation}
\label{eq:Bell_factorization}
p(\alpha,\beta|a,b)
=
\int d\lambda\,
q(\lambda)\,
p(\alpha|a,\lambda)\,
p(\beta|b,\lambda).
\end{equation}
Here, one also assumes measurement independence,
namely, that the settings are statistically
independent of the hidden variables characterizing the source. Eq.~\eqref{eq:Bell_factorization} expresses the idea that, once the relevant common
causes are specified by \(\lambda\), Alice's outcome probabilities depend only
on her local setting and likewise for Bob. The distribution \(q(\lambda)\)
need not represent merely an experimental lack of knowledge; it may equally
describe objective stochasticity in the underlying model.

Bell inequalities constrain the correlations compatible with 
factorization \eqref{eq:Bell_factorization}. Their violation shows that the
observed statistics cannot be reproduced by any locally realistic model, assuming
measurement independence and the closure of the relevant experimental
loopholes \cite{Hensen2015,Giustina2015,Shalm2015}.
The simplest example is the Clauser--Horne--Shimony--Holt (CHSH) inequality
\cite{CHSH}. For binary settings \(a,b\in\{0,1\}\) and binary outcomes
\(\alpha,\beta\in\{-1,1\}\), define 
\begin{equation}
E_{ab}
 =
\sum_{\alpha,\beta}
\alpha\beta\,p(\alpha,\beta|a,b).
\end{equation}
Every locally realistic correlation satisfies
\begin{equation}
\label{eq:CHSH}
|S|
 =
\left|
E_{00}+E_{01}+E_{10}-E_{11}
\right|
\leq 2.
\end{equation}
Quantum theory permits correlations exceeding this bound, up to the
Tsirelson limit \cite{Tsirelson} given by
\begin{equation}
|S|_{\mathrm{Quantum}}\leq 2\sqrt{2}.
\end{equation}

The Bell nonlocality should be distinguished from the quantum entanglement. The latter is necessary for a quantum
Bell violation, but not every entangled state violates a given Bell
inequality, and a violation generally requires appropriately chosen local
measurements \cite{Brunner2014}.
It is equally important to distinguish Bell nonlocality from superluminal
signaling and from nonlocality at the level of quantum field operators, i.e., microcausality. Quantum
correlations violating the Bell inequality remain compatible with the
no-signaling conditions
\begin{equation}
p(\alpha|a,b)=p(\alpha|a),
\qquad
p(\beta|a,b)=p(\beta|b).
\end{equation}
Also, in a local relativistic quantum field theory (QFT), microcausality ensures the compatibility of spacelike separated operations,
whereas Bell locality concerns whether their joint statistics admit the decomposition \eqref{eq:Bell_factorization}. Indeed,
Bell inequality violations can occur between spacelike separated local
observables even within a local relativistic QFT that satisfies microcausality
\cite{SummersWerner1985,SummersWerner1987}.

In the present work, we investigate how Bell-nonlocal correlations can emerge
from a dressing of local relativistic QFT field operators induced by noncommutative (NC) spacetime geometry. A NC spacetime is an algebraic structure described by coordinate operators $\hat{x}^{\mu}$ that satisfy some non-trivial commutation relations, the simplest of which are the canonical $\theta$-constant commutation relations, $[\hat{x}^{\mu},\hat{x}^{\nu}]=i\theta^{\mu\nu}$, where the NC deformation parameters $\theta^{\mu\nu}$ comprise a constant antisymmetric matrix \cite{ChaichianEtAl2004, ChaichianPresnajder2005, Aschieri2009Noncommutative}. The spacetime uncertainty relations  $\Delta\hat{x}^{\mu}\Delta\hat{x}^{\nu}\gtrsim \ell^{2}_{\text{NC}}$ encode the impossibility to sharply distinguish individual spacetime points \cite{Doplicher1994}. This hypothetical quantum structure that effectively emerges, most notably, in string theory \cite{Seiberg1999String} is suppose to replace the ordinary manifold structure of classical spacetime at some characteristic length scale $\ell_{\text{NC}}$, not necessarily Planck's. 
Working with free scalar QFT on the Moyal NC plane, defined by an Abelian Drinfel'd twist deformation \cite{Drinfeld1990, Aschieri2023Noncommutative, AschieriEtAl2005}, and following the twisted-statistics construction reviewed in \cite{Akofor2008}, we ask whether the underlying  NC spacetime structure can
supply a microscopic preparation mechanism for Bell-nonlocal correlations. 

The same Drinfel'd twist that defines a NC algebra of functions encoded by the $\star$-product also deforms the exchange structure of the multiparticle states \cite{FioreSchupp1996,Oeckl2000, BalachandranEtAl2006Spin, BalachandranEtAl2007Statistics}. 
For a constant Moyal deformation, the trace property of the $\star$-product
implies that the quadratic scalar action coincides with its classical
counterpart. Consequently, the propagator, dispersion
relation, free Hamiltonian, and one-particle states remain
unchanged. Nevertheless, the deformation survives in the tensor-product
structure of the theory: the twisted coproduct induces modified
multiparticle statistics, represented in Fock space by dressed creation
and annihilation operators \cite{ BalachandranPinzulQureshi2008,Akofor2008, AkoforEtAl2007}. Thus, the free propagation of each excitation
is ordinary, while the ordered composition of two or more excitations
carries momentum-dependent twist phases.

A phase attached to one fixed momentum pair is global and cannot affect an observable probability. The situation changes when an external source coherently prepares a quantum superposition of several different momentum-pair configurations. The dressing phases then become relative phases. Most can be removed by independent local basis changes, but one combination is invariant under all local rephasings. This
invariant phase acts as a controlled phase between two momentum-mode qubits
and can generate entanglement from a state that is separable in the
commutative limit ($\theta^{\mu\nu}\rightarrow 0$). The free field cannot dynamically create this entanglement; it comes from the preparation
interaction that we model by a weak external source coupled to the dressed free field. Once the NC-entangled state is prepared and distributed to Alice and Bob, they can apply suitable local measurements to test the correlations. Our results show that a violation of an appropriate CHSH Bell inequality serves as a signal of the purported underlying NC structure of spacetime. For an interesting work on observable effects of twisted statistics see \cite{Khan2008TwistedStatistics}.  

The paper is organized as follows.  Section~\ref{sec:gm} summarizes the Moyal
twist, twisted statistics, and the dressed scalar field. In
Section~\ref{sec:common_source_preparation} we derive the twist-entangled two-particle state prepared by a weak
source.  Section~\ref{sec:bell} formulates the effective two-qubit CHSH
protocol and derives the Bell parameter. We conclude in Section~\ref{sec:conclusion}. Technical details regarding twisted statistics and some CHSH calculations are provided in Appendices \ref{app:twisted_statistics} and \ref{app:chsh}.

\section{Moyal twist and field dressing}
\label{sec:gm}

\subsection{Star product and twisted statistics}

In Minkowski space, we can take any globally-inertial coordinate system $x^{\mu}$ and 
define a NC Moyal $\star$-product between functions on spacetime as
\begin{equation}
 f\star g
 =f\exp\left(
 \frac{\ii}{2}\overleftarrow{\partial}_\mu
 \theta^{\mu\nu}\overrightarrow{\partial}_\nu
 \right)g.
\end{equation}
For plane waves
\(e_p(x)=e^{-\ii p\cdot x}\),
\begin{equation}
 e_p\star e_q
 =e^{-\frac{\ii}{2}p\wedge q}e_{p+q},
 \qquad
 p\wedge q:=p_\mu\theta^{\mu\nu}q_\nu.
\end{equation}
Generally, for a set of mutually commuting vector fields $\{ X_{I}\}$, unrelated to a particular coordinate system, the corresponding $\star$-product is defined by a twist operator
\begin{equation}
    \mathcal{F}_{\theta}
    =
    \exp\left[
        -\frac{i}{2}\theta^{IJ}
        X_I\otimes X_J
    \right],
\end{equation}
with constant deformation parameters $\theta^{IJ}=-\theta^{JI}$. The corresponding $\star$-product is
\begin{equation}
    f\star g
    =
m_{0}\circ\mathcal{F}_{\theta}^{-1}(f\otimes g)=m_{\theta}(f\otimes g),
\end{equation}
where $m_{0}(f\otimes g)=fg$ is the multiplication map.

As explained in Appendix \ref{app:twisted_statistics}, the requirement of compatibility of the Poincar\'e transformations with the Moyal product leads to twisted Bose and Fermi statistics. For a complete account, see the review
\cite{Akofor2008} and references therein.
In particular, for momentum eigenstates, the twisted flip acts as
\begin{equation}
    \tau_{\theta}
    \left(
        |p\rangle\otimes|q\rangle
    \right)
    =
    e^{-ip\wedge q}
    |q\rangle\otimes|p\rangle.
    \label{app:twisted_flip_momentum}
\end{equation}
Thus, exchange is accompanied by a momentum-dependent phase. For example,
a twisted bosonic two-particle state may be written as
\begin{equation}
    |p,q\rangle_{\theta,+}
    =
    \frac{1}{\sqrt{2}}
    \left(
        |p\rangle\otimes|q\rangle
        +
        e^{-ip\wedge q}
        |q\rangle\otimes|p\rangle
    \right),
    \label{app:twisted_bosonic_state}
\end{equation}
for distinct orthogonal one-particle states.
The vacuum and the one-particle sector remain unchanged. The deformation resides in the
multiparticle exchange structure and in the relative phases between
different multiparticle momentum configurations. 

\subsection{Dressing transformation}

The bare free real scalar field is expanded in terms of bare creation and anihilation operators as
\begin{equation}
 \label{eq:dressed_field_modes}
 \hat{\phi}_0(x)
 =
 \int d\mu(p)
 \left[
 b_p e^{-\ii p\cdot x}
 +b_p^\dagger e^{\ii p\cdot x}
 \right],
\end{equation}
where
\begin{equation}
 d\mu(p)
 =\frac{d^3\bm p}{(2\pi)^{3}\;2 E_{\bm p}},
 \qquad
 E_{\bm p}=\sqrt{\bm p^2+m^2}.
\end{equation}
The NC-dressed oscillators can be represented on the ordinary bosonic Fock space
by the dressing transformation
\begin{align}
\label{eq:dressing_annihilation}
 d_p&=b_p\exp\left(-\frac{\ii}{2}p\wedge P\right),\\
\label{eq:dressing_creation}
 d_p^\dagger&=b_p^\dagger
 \exp\left(+\frac{\ii}{2}p\wedge P\right), 
\end{align}
with the total momentum operator
\begin{equation}\label{P_commutator}
 P_\mu=\int d\mu(k)\,k_\mu b_k^\dagger b_k.
\end{equation}
Note that since
\begin{equation}\label{P_commutator}
 [P_\mu,b_p^\dagger]=p_\mu b_p^\dagger,
\end{equation}
we have $[p\wedge P,b^{\dagger}_{p}]=0$, justifying the definition of the dressed creation operator (\ref{eq:dressing_creation}).

While bare mode-operators $b_p$ and $b^\dagger_q$ satisfy ordinary bosonic algebra, an exchange of dressed mode-operators picks up a momentum-dependent NC phase,
\begin{equation}
\label{eq:twisted_exchange}
 d_p^\dagger d_q^\dagger
 =e^{\ii p\wedge q}d_q^\dagger d_p^\dagger.
\end{equation}
Acting on the translation-invariant vacuum,
\begin{equation}
 P_\mu\ketvac=0,
\end{equation}
gives unchanged one-particle states $    d_q^\dagger\ket0
    =
    b_q^\dagger\ket0$, while 
\begin{equation}
 \label{eq:pair_phase}
 d_p^\dagger d_q^\dagger\ketvac
 =e^{\frac{\ii}{2}p\wedge q}
 b_p^\dagger b_q^\dagger\ketvac.
\end{equation}
For one fixed pair \((p,q)\), the factor in Eq.~\eqref{eq:pair_phase} is only
a global phase. It becomes a relative phase only in a coherent superposition
of different momentum pairs.

A simple substitution yields a NC-dressed field 
\begin{equation}
 \label{eq:dressed_field}
 \hat{\phi}_\theta(x)
 =\hat{\phi}_0(x)
 \exp\left(
 \frac12\overleftarrow\partial_\mu
 \theta^{\mu\nu}P_\nu
 \right),
\end{equation}
where $\partial_{\mu}$ acts to the left on the bare field $\hat{\phi}_{0}$ and the momentum operator $P_{\nu}$ acts on the Fock space to the right. 
The dressing is not an additional deformation independent of the $\star$-
product. It is the Fock-space realization of the same NC twist
\cite{Akofor2008}. It is important to note that the dressing factor contains the total
momentum operator \(P_\nu\). Consequently, the action of a dressed field operator depends on the total-momentum
of the state on which it acts. In this sense, the dressed field is nonlocal operator - although it is a fixed operator-valued
field, its action is sensitive to the momentum content of the entire
multiparticle state. This feature is irrelevant in the vacuum and in
the one-particle sector, but produces momentum-dependent phases when
successive excitations are created. 

The normal-ordered free Hamiltonian is
\begin{equation}
 H_0
 =\int d\mu(p)\,E_{\bm p}\,b_p^\dagger b_p.
\end{equation}
Because \(p\wedge p=0\), the dressing factors cancel in the number operator, $d_p^\dagger d_p=b_p^\dagger b_p$, and the Hamiltonian remains the same. 
The relevant NC information entirely lies in the kinematic exchange structure of the multiparticle momentum 
states. As we shall see in the following, we can use this twisted statistics 
as a resource that can be tested by the Bell protocol.

\section{State preparation with a weak external source}
\label{sec:common_source_preparation}

We consider a standard common-source geometry in which the quantum state
is prepared in the joint causal past of two spacelike-separated
laboratories. Let \(\mathcal O_A\) and \(\mathcal O_B\) denote Alice's and
Bob's measurement regions, respectively, with
\(\mathcal O_A\) spacelike separated from \(\mathcal O_B\). The source is
localized in a compact region \(\mathcal O_S\) satisfying
\begin{equation}
    \mathcal O_S
    \subset
    J^{-}(\mathcal O_A)
    \cap
    J^{-}(\mathcal O_B).
    \label{eq:common_past_geometry}
\end{equation}
where $J^-$ stands for causal past. 
All interactions responsible for the preparation therefore take place
before the measurement settings are selected and before the
spacelike-separated measurements are performed.

A major hypothesis is that the classical source $J(x)$ couples locally and linearly to the dressed quantum field, yielding an interaction picture evolution
\begin{equation}
    U_{J}
    =\mathcal{T}\exp\left(-i
    \lambda
    \int d^4x\,
    J(x)\hat{\phi}_\theta(x)\right),
\end{equation}
where $\lambda$ is a coupling constant. This interaction creates a one-particle excitation in the first order in \(\lambda\).

Conceptually, the source should prepare two outgoing wave-packets: the \(A\)-packet propagates toward
Alice's laboratory $\mathcal{O}_{A}$ and the \(B\)-packet toward Bob's $\mathcal{O}_{B}$. 
We assume that the source activity consists of two temporally separated pulses,
supported in regions
\(\mathcal O_S^{(A)},\mathcal O_S^{(B)}\subset\mathcal O_S\), with $\mathcal O_S^{(B)}$ being in the chronological past of $\mathcal O_S^{(A)}$, namely
\begin{equation}
\mathcal O_S^{(B)}\subset
    I^{-}\!\left(\mathcal O_S^{(A)}\right).
    \label{eq:source_pulse_order}
\end{equation}

\begin{figure}[h!]
    \centering
\begin{tikzpicture}[scale=0.62,transform shape, thick]

    \draw[->] (-7,0) -- (6.72,0) node[right] {$x$};
    \draw[->] (0,-2.1) -- (0,4.4) node[above] {$t$};

    \coordinate (LB) at (-4,0);
    \coordinate (RB) at (4,0);
    \coordinate (C)  at (0,-3.2);

    \fill[orange!15] (-3,3) ellipse (0.65 and 0.38);
    \draw[orange!80!black] (-3,3) ellipse (0.65 and 0.38);
    \node at (-3,3) {$\mathcal{O}_A$};
    \fill[green!15] (3,3) ellipse (0.65 and 0.38);
    \draw[green!60!black] (3,3) ellipse (0.65 and 0.38);
    \node at (3,3) {$\mathcal{O}_B$};

    \draw[dashed, orange!80!black]
        (-7,-0.35) -- (-3.65,3);
    \draw[dashed, orange!80!black]
        (-2.35,3) -- (3.65,-3);
   \begin{scope}
    \fill[orange!30, fill opacity=0.25]
        (-3.65,3)
        -- (-7,-0.35)
        -- (-7,-3)
        -- (3.65,-3)
        -- (-2.35,3)
        -- cycle;
\end{scope}
    \draw[dashed, green!60!black]
        (-3.65,-3) -- (2.35,3) ;
    \draw[dashed, green!60!black]
        (3.65,3) -- (7,-0.35) ;
        \begin{scope}
    \fill[green!30, fill opacity=0.25]
        (2.35,3)
        -- (-3.65,-3)
        -- (7,-3)
        -- (7,-0.35)
        -- (3.65,3)
        -- cycle;
\end{scope}


    \fill[red!15] (0,-0.2) ellipse (0.65 and 0.38);
    \draw[red!80!black] (0,-0.2) ellipse (0.65 and 0.38);
    \node at (0,-0.2) {$\mathcal{O}_S^{(A)}$};
   \draw[red!60!black,line width=0.5pt]
    (-0.65,-0.2) -- (-2.83,2.63);

\begin{scope}[shift={(-1.74,1.215)}]
    \draw[
        rotate=127.61,
        red!80!black,
        line width=0.5pt,
        domain=-0.45:0.45,
        samples=120,
        smooth
    ]
    plot (\x,{0.15*exp(-12*\x*\x)*sin(58*\x r)});
\end{scope}

\begin{scope}[shift={(-1.74,1.215)}]
    \draw[
        rotate=127.61,
        red!80!black,
        line width=0.5pt,
        domain=-0.45:0.45,
        samples=120,
        smooth
    ]
    plot (\x,{0.5*exp(-12*\x*\x)*sin(12*\x r)});
\end{scope}

    \fill[blue!15] (0,-1) ellipse (0.65 and 0.38);
    \draw[blue!80!black] (0,-1) ellipse (0.65 and 0.38);
    \node at (0,-1) {$\mathcal{O}_S^{(B)}$};

  \draw[blue!60!black,line width=0.5pt]
    (0.65,-1) -- (2.65,2.67);

\begin{scope}[shift={(1.65,0.835)}]
    \draw[
        rotate=61.41,
        blue!80!black,
        line width=0.5pt,
        domain=-0.45:0.45,
        samples=120,
        smooth
    ]
    plot (\x,{0.15*exp(-12*\x*\x)*sin(58*\x r)});
\end{scope}

\begin{scope}[shift={(1.65,0.835)}]
    \draw[
        rotate=61.41,
        blue!80!black,
        line width=0.5pt,
        domain=-0.45:0.45,
        samples=120,
        smooth
    ]
    plot (\x,{0.5*exp(-12*\x*\x)*sin(12*\x r)});
\end{scope}

     \node at (0,-2.5) {$J^{-}(\mathcal{O}_A)\cap J^{-}(\mathcal{O}_B)$};
   
\end{tikzpicture}

 \caption{A schematic representation of the relativistic Bell protocol. Alice and Bob reside in their space-like separated laboratories $\mathcal{O}_{A}$ and $\mathcal{O}_{B}$, respectively. The source coupled to the dressed quantum field generates  excitations in time-ordered preparation regions $\mathcal{O}^{(B)}_{S}$ and $\mathcal{O}^{(B)}_{S}$ belonging to the common causal past of $\mathcal{O}_{A}$ and $\mathcal{O}_{B}$. The excitations are arranged to propagate towards $\mathcal{O}_{B}$ and $\mathcal{O}_{A}$, respectively. Each signal is a single-particle excitation in a state of quantum superposition of two orthogonal wave-packets.}
\end{figure}
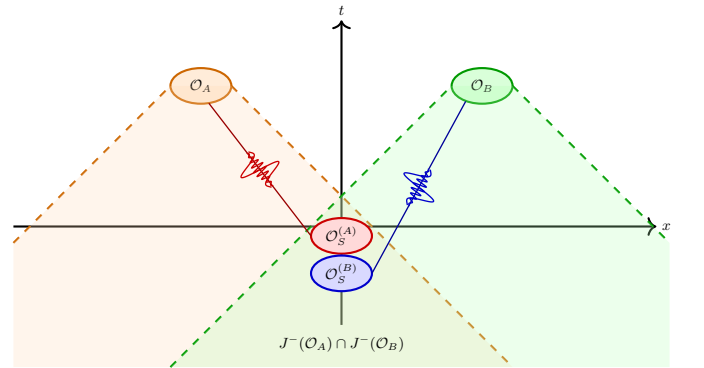
Because the two source pulses have Lorentz-invariant time-ordering as in
Eq.~\eqref{eq:source_pulse_order}, the pulse producing the Bob-directed packet is completed before
the pulse producing the Alice-directed packet begins, and we have the following  evolution
factorization   
\begin{align}
    &U_J
    =\mathcal{T}\exp\Bigg(
    -i\lambda_A
    \int_{\mathcal O_S^{(A)}}d^4x\,
    J_A(x)\hat{\phi}_\theta(x)\Bigg)
   \nonumber\\
   \times&\mathcal{T}
    \exp\Bigg(-i\lambda_B
    \int_{\mathcal O_S^{(B)}}d^4y\,
J_B(y)\hat{\phi}_\theta(y)\Bigg)=U_AU_B.
\label{eq:linear_common_source}
\end{align}
where \(U_B\) describes the earlier Bob-directed pulse and \(U_A\) the
later Alice-directed pulse.
Here, \(J_A\) and \(J_B\) are classical source profiles, while
\(\lambda_A\) and \(\lambda_B\) control the strengths of the two
pulses. Once the wave-packets are excited by the source, we can engineer their distribution to Alice and Bob in various ways, but they would still be governed by free field dynamics. 

The dressed field is decomposed as
\begin{equation}
    \hat{\phi}_\theta(x)
    =
    \hat{\phi}_\theta^{(+)}(x)
    +
    \hat{\phi}_\theta^{(-)}(x),
\end{equation}
where the creation part is
\begin{equation}
    \hat{\phi}_\theta^{(-)}(x)
    =
    \int d\mu(p)\,
    d_p^\dagger e^{\ii p\cdot x}.
\label{eq:dressed_creation_field}
\end{equation}
The two source profiles define the momentum-space wave packets
\begin{align}
    f_A(p)
    &=
    \int_{\mathcal O_S^{(A)}}d^4x\,
    J_A(x)e^{\ii p\cdot x},
    \\
    f_B(q)
    &=
    \int_{\mathcal O_S^{(B)}}d^4y\,
    J_B(y)e^{\ii q\cdot y}.
    \label{eq:source_wavepacket_profiles}
\end{align}
The corresponding dressed creation operators are
\begin{align}
    A_\theta^\dagger[f_A]
    &:=
    \int d\mu(p)\,
    f_A(p)d_p^\dagger,
    \\
    B_\theta^\dagger[f_B]
    &:=
    \int d\mu(q)\,
    f_B(q)d_q^\dagger.
    \label{eq:source_packet_operators}
\end{align}

Expanding to second order in the linear
couplings and acting on the free field theory vacuum gives
\begin{align}
    &U_J\ket0
    =
    \ket0
    -
    \ii\lambda_A
    A_\theta^\dagger[f_A]\ket0
    -
    \ii\lambda_B
    B_\theta^\dagger[f_B]\ket0
    \nonumber\\
    &-
    \lambda_A\lambda_B
    A_\theta^\dagger[f_A]
    B_\theta^\dagger[f_B]\ket0
    \nonumber\\
    &-
    \frac{\lambda_A^2}{2}
    \left(
        A_\theta^\dagger[f_A]
    \right)^2
    \ket0
    -
    \frac{\lambda_B^2}{2}
    \left(
        B_\theta^\dagger[f_B]
    \right)^2
    \ket0
    +
    \mathcal O(\lambda^3).
    \label{eq:linear_source_expansion}
\end{align}
Eq.~\eqref{eq:linear_source_expansion} exhibits an important
feature of a linear source. It prepares a coherent superposition
of the vacuum, one-particle states, two-particle states, and
higher-particle sectors. In particular, the desired contribution
containing one outgoing excitation in each channel appears at second
order,
\begin{equation}
    \ket{\psi_{AB}^{(2)}}
    =
    -
    \lambda_A\lambda_B
    A_\theta^\dagger[f_A]
    B_\theta^\dagger[f_B]\ket0.
    \label{eq:second_order_AB_state}
\end{equation}

Let \(\Pi_{1_A,1_B}\) denote the projector onto the sector containing
one excitation in the \(A\)-channel and one excitation in the
\(B\)-channel. Then
\begin{equation}
    \Pi_{1_A,1_B}U_J\ket0
    =
    -
    \lambda_A\lambda_B
    A_\theta^\dagger[f_A]
    B_\theta^\dagger[f_B]\ket0
    +
    \mathcal O(\lambda^3).
    \label{eq:projected_source_state}
\end{equation}
The normalized conditional two-particle state is therefore
\begin{equation}
    \ket{\Psi_{AB,\theta}}
    =\mathcal{N}
    A_\theta^\dagger[f_A]
    B_\theta^\dagger[f_B]\ket0.
\label{eq:conditional_common_source_state}
\end{equation}
The coupling constants determine the probability of obtaining this
sector, but do not enter the normalized conditional state. 


The projection in Eq.~\eqref{eq:projected_source_state} should be distinguished from setting-dependent postselection. For theoretical analysis, one may work directly in the sector with one excitation in each outgoing mode
family. However, in an operational Bell experiment, a 
successful preparation event should be identified before Alice and Bob
select their measurement settings. Alternatively, vacuum and no-detection events must be retained among the
measurement outcomes.

The NC phase becomes transparent after expressing the
dressed creation operators in terms of bare creation operators and rewriting the state~\eqref{eq:conditional_common_source_state} in the ordinary Fock basis as
\begin{equation}
    \ket{\Psi_{AB,\theta}}
    =\mathcal{N}
    \int d\mu(p)d\mu(q)\,
    f_A(p)f_B(q)
    e^{\frac{\ii}{2}p\wedge q}
    b_p^\dagger b_q^\dagger\ket0.
    \label{eq:common_source_ordinary_basis}
\end{equation}
Therefore, the earlier
pulse prepares the Bob-directed one-particle packet and the later pulse
prepares the Alice-directed packet, but its dressed creation operator
acts on a state that already carries the momentum of the first
excitation. This produces the joint phase
\(e^{\frac{i}{2}p\wedge q}\). The individual one-particle states remain
undeformed; the NC structure is encoded in their joint two-particle
amplitude.

To construct a two-qubit state, each outgoing particle is prepared in a
coherent superposition of two orthonormal wave-packet modes:
\begin{align}
    f_A
    &=
    \alpha_0f_0^A
    +
    \alpha_1f_1^A,
    \\
    f_B
    &=
    \beta_0f_0^B
    +
    \beta_1f_1^B,
    \label{eq:logical_source_profiles}
\end{align}
where
\begin{align}
    \int d\mu(p)\,
    f_i^{A*}(p)f_k^A(p)
    &=
    \delta_{ik},
    \\
    \int d\mu(q)\,
    f_j^{B*}(q)f_l^B(q)
    &=
    \delta_{jl}.
\end{align}
The corresponding packet-creation operators are
\begin{align}
    A_{\theta,i}^\dagger
    &:=
    A_\theta^\dagger[f_i^A],
    \\
    B_{\theta,j}^\dagger
    &:=
    B_\theta^\dagger[f_j^B].
\end{align}
The conditional state becomes
\begin{equation}
    \ket{\Psi_{AB,\theta}}
    =\mathcal{N}
    \sum_{i,j=0}^{1}
    \alpha_i\beta_j
    A_{\theta,i}^\dagger
    B_{\theta,j}^\dagger\ket0.
    \label{eq:logical_source_state_dressed}
\end{equation}
For sufficiently narrow wave-packets centered around momenta \(p_i\) and
\(q_j\), the twist phase is approximately constant over each pair of
momentum supports. One then approximately has
\begin{equation}
    A_{\theta,i}^\dagger
    B_{\theta,j}^\dagger\ket0
    \simeq
    e^{\frac{\ii}{2}p_i\wedge q_j}
    A_{0,i}^\dagger
    B_{0,j}^\dagger\ket0.
    \label{eq:narrow_packet_source_approximation}
\end{equation}
Defining the computational  basis states
\begin{equation}
    \ket{ij}
    :=
    A_{0,i}^\dagger
    B_{0,j}^\dagger\ket0,
\end{equation}
the two-qubit state takes the form
\begin{equation}
    \ket{\Psi_\theta}
    =\mathcal{N}
    \sum_{i,j=0}^{1}
    \alpha_i\beta_j
    e^{\ii\phi_{ij}}
    \ket{ij},
    \qquad
    \phi_{ij}
    =
    \frac12p_i\wedge q_j.
    \label{eq:logical_common_source_state}
\end{equation}
In the commutative limit ($\theta^{\mu\nu}\rightarrow 0$), all twist phases vanish and the state
factorizes as
\begin{equation}
    \ket{\Psi_0}
    =\mathcal{N}
    \left(
        \alpha_0\ket0_A
        +
        \alpha_1\ket1_A
    \right)
    \otimes
    \left(
        \beta_0\ket0_B
        +
        \beta_1\ket1_B
    \right).
    \label{eq:commutative_source_product_state}
\end{equation}
So, in the commutative limit, the source would produce a simple product state. 
For nonzero \(\theta^{\mu\nu}\), the four phases in
Eq.~\eqref{eq:logical_common_source_state} cannot, in general, be
removed by independent rephasings of Alice's and Bob's local bases. The
local-unitary invariant combination is
\begin{align}
    \chi_\theta
    &:=
    \phi_{00}
    +
    \phi_{11}
    -
    \phi_{01}
    -
    \phi_{10}
    \nonumber\\
    &=
    \frac12
    (p_0-p_1)_\mu
    \theta^{\mu\nu}
    (q_0-q_1)_\nu.
\label{eq:common_source_invariant_phase}
\end{align}
This is the nonlocal phase responsible for the entanglement of the
computational  wave-packet modes. We can therefore work with the equivalent state
\begin{equation}
    \vert\Psi_{\theta}\rangle\sim\ket{\Psi_\chi}
    =
    \frac{1}{2}
    \left(
        \ket{00}
        +
        \ket{01}
        +
        \ket{10}
        +
        e^{\ii\chi_\theta}\ket{11}
    \right).
\end{equation}


The two computational  modes propagating toward each laboratory are narrow wave
packets centered around distinct momenta - $p_0$ and $p_1$ for Alice and $q_0$ and $q_1$ for Bob. Any difference in their group
velocities and arrival-time distributions is a matter of source and
channel engineering. We assume that the preparation is arranged so that
the two modes are coherently accessible within the same local measurement
window.

The fundamental preparation procedure is therefore entirely local and
causal. A localized source in the common past of the two space-like separated laboratories couples linearly to the
dressed field through two sequential pulses. The desired state occurs
in the second-order sector containing one excitation in each outgoing
channel. After conditioning on this sector, the ordered action of the
dressed creation operators produces the momentum-dependent NC phase that
entangles the two computational  wave-packet degrees of freedom.

\section{Bell--CHSH protocol}
\label{sec:bell}

\subsection{Local wave-packet observables}
\label{sec:local_mode_observables}

Alice and Bob perform measurements on the two computational  wave-packet modes
introduced in the previous section. On Alice's
one-excitation subspace, the direct which-mode observable is
\begin{equation}
    Z_A
    =
    A_{0,0}^\dagger A_{0,0}
    -
    A_{0,1}^\dagger A_{0,1},
    \label{eq:ZA}
\end{equation}
while the complementary mode-mixing observable is
\begin{equation}
    X_A
    =
    A_{0,0}^\dagger A_{0,1}
    +
    A_{0,1}^\dagger A_{0,0}.
    \label{eq:XA}
\end{equation}
Restricted to the computational  one-particle subspace, these operators act as
the Pauli operators,
\begin{align}
    Z_A\ket0_A
    &=
    \ket0_A,
    &
    Z_A\ket1_A
    &=
    -\ket1_A,
    \\
    X_A\ket0_A
    &=
    \ket1_A,
    &
    X_A\ket1_A
    &=
    \ket0_A.
\end{align}
Bob's observables are defined analogously,
\begin{align}
    Z_B
    &=
    B_{0,0}^\dagger B_{0,0}
    -
    B_{0,1}^\dagger B_{0,1},
    \\
    X_B
    &=
    B_{0,0}^\dagger B_{0,1}
    +
    B_{0,1}^\dagger B_{0,0}.
\end{align}

Operationally, the \(Z\) measurement distinguishes the two local
wave-packet modes. An \(X\) measurement is implemented by first applying
a balanced mixer between the two modes and then performing the same
which-mode readout. More generally, the two possible observables chosen by Alice and Bob
in the local \(XZ\)-planes are
\begin{align}
    A_a
    &=
    \cos\vartheta_a\,Z_A
    +
    \sin\vartheta_a\,X_A,
    \qquad
    a\in\{0,1\},
    \\
    B_b
    &=
    \cos\varphi_b\,Z_B
    +
    \sin\varphi_b\,X_B,
    \qquad
    b\in\{0,1\}.
    \label{eq:local_XZ_observables}
\end{align}
Alice and Bob measure two
distinct outgoing wave-packet modes in their respective laboratories.
The overlap between the A and B-modes is negligible, $\langle f_i^A|f_j^B\rangle \simeq 0$, so that they can be treated as
independent effective subsystems. 
The measurement operators therefore take the bipartite form
\begin{equation}
    \mathcal{A}_a:
    =
    A_a\otimes\mathrm{1}_B,
    \qquad
    \mathcal{B}_b:
    =
    \mathrm{1}_A\otimes B_b.
\end{equation}
Consequently,
\begin{equation}
[\mathcal{A}_a,\mathcal{B}_b]=0
\end{equation}
for all settings \(a,b\in\{0,1\}\), as required for independently
controlled measurements on the two Bell wings.

The local wave-packet basis is fixed by the calibration of the mode
mixers and detectors and is not twist-deformed together with the
prepared state. The NC deformation is therefore encoded in
\(\ket{\Psi_\theta}\) relative to this fixed measurement basis. 

\subsection{CHSH violation}
\label{sec:CHSH_violation}

For settings \(a,b\in\{0,1\}\) and binary outcomes
\(\alpha,\beta\in\{\pm1\}\), we define the correlators
\begin{equation}
    E(a,b)
    =
    \bra{\Psi_\theta}
    A_aB_b
    \ket{\Psi_\theta}
    =
    \sum_{\alpha,\beta=\pm1}
    \alpha\beta\,
    P(\alpha,\beta|a,b),
\end{equation}
and the CHSH parameter
\begin{equation}
    S_\theta
    =
    E(0,0)
    +
    E(0,1)
    +
    E(1,0)
    -
    E(1,1).
    \label{eq:CHSH_parameter}
\end{equation}

As shown explicitly in
Appendix~\ref{app:chsh}, the settings
\eqref{eq:optimal_settings} maximize the CHSH parameter and give
\begin{align}
    S_\theta^{\max}
    &=
    2\sqrt{1+C_\theta^2}
    \nonumber\\
    &=
    2\sqrt{
        1+
        \sin^2\left(
            \frac{\chi_\theta}{2}
        \right)
    }.
    \label{eq:Smax}
\end{align}
This agrees with the general maximal-CHSH result for pure two-qubit
states \cite{Horodecki1995}.
Consequently, if $
    \chi_\theta\neq0
    \pmod{2\pi}$, then $
    C_\theta>0$ and so $
    S_\theta^{\max}>2$.
Thus, every entangled state generated by a nontrivial controlled NC phase in
this family violates the CHSH inequality for a suitable choice of local
mode measurements.
For $\chi_\theta
    =
    \pi
    \pmod{2\pi}$,
the concurrence reaches \(C_\theta=1\), the state is maximally
entangled, and the maximal quantum value $S_\theta^{\max}
    =
    2\sqrt2$ is obtained.

For weak noncommutativity, 
the concurrence and maximal CHSH value behave as
\begin{align}
    C_\theta
    &=
    \frac{|\chi_\theta|}{2}
    +
    \mathcal O(\theta^3),
    \\
    S_\theta^{\max}
    &=
    2
    +
    \frac{\chi_\theta^2}{4}
    +
    \mathcal O(\theta^4).
    \label{eq:smalltheta_CHSH}
\end{align}
The entanglement generated by the twisted multiparticle structure  therefore appears linearly in the NC deformation parameters,
whereas the optimized excess above the CHSH bound is
quadratic.

\subsection{Momentum-space geometry}
\label{subsec:momentum_geometry}

The NC phase depends not only on the magnitude of the deformation
parameter, but also on the geometry of the selected momentum modes. Recall that
\begin{equation}
    \chi_\theta
    =
    \frac{1}{2}
    (p_0-p_1)_\mu
    \theta^{\mu\nu}
    (q_0-q_1)_\nu .
\end{equation}
Introducing the local momentum separations
\begin{equation}
    \Delta p:=p_0-p_1,
    \qquad
    \Delta q:=q_0-q_1,
\end{equation}
this can be written compactly as
\begin{equation}
    \chi_\theta
    =
    \frac{1}{2}\Delta p\wedge\Delta q.
\end{equation}
For purely spatial noncommutativity in the \((x,y)\)-plane, we take
\begin{equation}
    \theta^{12}=-\theta^{21}=\theta,
    \qquad
    \theta^{0i}=0,
\end{equation}
with all other components vanishing. The noncommutative phase then becomes
\begin{equation}
    \chi_\theta
    =
    \frac{\theta}{2}
    \left(
    \Delta p_x\Delta q_y
    -
    \Delta p_y\Delta q_x
    \right)=
    \frac{\theta}{2}
    \left(
    \Delta\bm p\times\Delta\bm q
    \right)_z.
\end{equation}
The phase is therefore proportional to the oriented area of the parallelogram
spanned by the two local momentum separations in the NC plane. This
gives a direct geometrical interpretation of the deformation: the relevant
quantity is not an individual momentum, but the relative orientation of the
momentum-mode splittings used to define Alice's and Bob's effective qubits.

In particular, if the two momentum separations are collinear, $
    \Delta\bm p\parallel\Delta\bm q$,
then $
    \chi_\theta=0$.
The dressed state is then locally equivalent to the commutative product state,
and no Bell--CHSH violation is generated by the NC dressing. More
generally, a one-dimensional arrangement of momentum modes cannot reveal purely
spatial noncommutativity, since it does not enclose a nonzero oriented area in
momentum space.

The effect is largest when the two momentum separations lie in the
NC plane and are mutually orthogonal. As a simple example, take
\begin{equation}
    \Delta\bm p
    =
    2k\,\widehat{\bm x},
    \qquad
    \Delta\bm q
    =
    2\ell\,\widehat{\bm y}.
\end{equation}
The invariant phase is then
\begin{equation}
    \chi_\theta
    =
    2\theta k\ell .
\end{equation}
and the maximal CHSH value 
\begin{equation}
        S_\theta
    =
    2\sqrt{
    1+\sin^2(\theta k\ell)
    }.
\end{equation}
Maximal entanglement is reached whenever
\begin{equation}
    \theta k\ell
    =
    \frac{\pi}{2}+n\pi,
    \qquad
    n\in\mathbb Z,
\end{equation}
for which $
    S_\theta=2\sqrt{2}$.
The NC contribution vanishes whenever
\begin{equation}
    \theta k\ell=n\pi,
\end{equation}
in which case $S_\theta=2$.

This momentum-space interpretation also clarifies the role of the common
preparation stage. The source must coherently populate at least two distinct
momentum modes in each outgoing sector, and the corresponding local mode
separations must probe a nonvanishing component of
\(\theta^{\mu\nu}\). The measurement apparatus itself may remain unchanged -
Alice and Bob only need to implement the local mode mixers defining their
\(X\)-type observables and the mode-resolved measurements defining their
\(Z\)-type observables. The NC information is encoded in the
geometry of the momentum configurations selected during preparation.

\section{Conclusion}
\label{sec:conclusion}

We have constructed a CHSH Bell protocol for a free real quantum scalar field on the NC
Moyal plane. The NC deformation enters
through twisted multiparticle statistics and its Fock-space dressing
representation. In using a model of local linear coupling between the external source and the quantum field, we postulated that the physical excitations are NC-dressed particles, i.e., that source couples to the dressed field.   
The source prepares an entangled state in the twisted two-particle sector, and
local mode
measurements yield a NC-dependent violation of the Bell CHSH inequality for momentum modes.  

This work was partly motivated by recent proposals in which
quantum spacetime structures affect the status of probabilities
themselves
\cite{DEsposito2026Indefinite,CastroRuiz2026Probabilities}.
In the quantum group construction of
\cite{DEsposito2026Indefinite}, the probabilities associated with
different measurements become noncommuting operators. In
\cite{CastroRuiz2026Probabilities}, measurements performed relative
to nonideal quantum reference frames lead to intrinsically indefinite
relational frequencies and expectation values, whose correlations can
be tested through Bell-type inequalities.

The present construction is in a sense complementary to these approaches. Here,
the Born probabilities remain ordinary numerical probabilities and the
local detector observables retain their standard form. The NC
structure instead modifies the structure of multiparticle
excitations, producing a NC-dependent entangled state before the
local measurements are performed. The resulting CHSH violation
therefore witnesses the twisted multiparticle
statistics rather than an incompatibility of probability observables.
Taken together, these approaches illustrate different operational
levels at which quantum spacetime structures may enter quantum theory:
through the probability and reference-frame sector, or through the
preparation and structure of multiparticle states.
The central modeling assumption that remains to be investigated
microscopically is the asymmetry between preparation and measurement -
the source is assumed to couple to the NC dressed field, whereas the local
wave-packet basis of the mode mixers and detectors is kept fixed. A
microscopic model should determine whether this state-measurement split
is dynamically realizable.

A further direction is to extend the construction beyond a fixed
Minkowski background. In a gravitational setting, the localization of
the source and detectors, as well as the definition of their local mode
bases, should ultimately be formulated relationally. The quantum
reference field framework developed in
\cite{Chen2026QuantumReferenceFields} provides a possible starting
point for this problem, since it describes local field-theoretic
reference systems, relational gauge-invariant observables, and
operational measurement schemes in linearized quantum gravity. It would
be interesting to investigate how twisted multiparticle states and
their Bell correlations are represented relative to such quantum
reference fields, and whether the NC phase retains an invariant
operational meaning under transformations between quantum field
perspectives. 

\section*{Acknowledgement}
\label{sec:acknowledgement}
This work is supported by the Faculty of Physics, University of Belgrade, through Grant No. 451-03-136/2025-03/200162 by the Ministry of Science, Technological Development and Innovations of the Republic of Serbia. Also, I.\Dj., A.G. and D.G. acknowledge the support of the Science Fund
of the Republic of Serbia, Grant No. 9029-YF-SAIGE, Twisted Holography: A Holographic Stance on the Quantum Superposition of
Spacetimes - HOLISTIQUS.  

\appendix

\section{Twisted statistics}
\label{app:twisted_statistics}

We introduce a noncommutative spacetime using an Abelian Drinfel'd twist,
\begin{equation}
    \mathcal{F}_{\theta}
    =
    \exp\left[
        -\frac{i}{2}\theta^{IJ}
        X_I\otimes X_J
    \right],
    \qquad
    [X_I,X_J]=0,
    \label{app:general_twist}
\end{equation}
where $\{X_I\, \vert\, I=1,\dots,s\leq D\}$ is a set of mutually commuting vector fields on the classical (generally curved) spacetime
manifold (note that the number of vector fields used to define the twist need not be equal to the number of spacetime dimensions $D$),
and the deformation parameters
\(\theta^{IJ}=-\theta^{JI}\) are constant. The corresponding
\(\star\)-product between functions is
\begin{equation}
    f\star g
    =
    m_0\circ\mathcal{F}_{\theta}^{-1}(f\otimes g)
    \equiv
    m_{\theta}(f\otimes g),
    \label{app:star_product}
\end{equation}
where
\begin{equation}
    m_0(f\otimes g)=fg
\end{equation}
is the ordinary pointwise multiplication map. If the vector fields \(X_I\) mutually commute, the \(\star\)-product is
associative, but it is
generically noncommutative.

To first order in the deformation parameters,
\begin{equation}
    f\star g
    =
    fg
    +
    \frac{i}{2}\theta^{IJ}
    \mathcal{L}_{X_I}f\,
    \mathcal{L}_{X_J}g
    +
    \mathcal{O}(\theta^2),
    \label{app:star_expansion}
\end{equation}
where
\begin{equation}
    \mathcal{L}_{X_I}f
    =
    X_I[f]
    =
    X_I^{\mu}\partial_{\mu}f.
\end{equation}
Applying the \(\star\)-product to the coordinate functions gives
\begin{equation}
    [x^{\mu},x^{\nu}]_{\star}
    =
    i\theta^{IJ}
    X_I^{\mu}(x)X_J^{\nu}(x)
    +
    \mathcal{O}(\theta^3)
    \equiv
    i\theta^{\mu\nu}(x).
    \label{app:coordinate_commutator}
\end{equation}
If the vector fields are chosen to be coordinate derivatives,
\(X_I=\partial/\partial x^\mu\) (at least in a patch of spacetime), in that particular coordinate system, the $\star$-commutator becomes constant, $[x^{\mu},x^{\nu}]_{\star}=i\theta^{IJ}\delta^{\mu}_{I}\delta^{\nu}_{J}=i\theta^{\mu\nu}$. 
In a generic coordinate system, unrelated to the vector fields defining the twist, the $\star$-commutator would not be constant. This does not represent
a change in the underlying deformation since the constants \(\theta^{IJ}\)
and the vector fields \(X_I\) continue to define the same
twist.

The set \(\{X_I\}\) is thus an additional geometric structure on the
spacetime manifold. Writing the twist in terms of vector fields makes
the construction covariant under coordinate changes, since the
components \(X_I^{\mu}\) transform as vector components. However, this coordinate
covariance should not be confused with invariance under an
arbitrary change in the vector fields \(X_I\). A specific choice of vector fields selects the NC structure being
considered.

\subsection{Moyal twist and Poincar\'e covariance}
\label{app:twisted_poincare}

For the remainder of this Appendix, we specialize in Minkowski spacetime
and adapt the twist vector fields to some globally inertial frame, $X_I=\partial_{\mu}$, which gives us the standard Moyal twist 
\begin{equation}
    \mathcal{F}_{\theta}
    =\exp\left[
        -\frac{i}{2}
        \theta^{\mu\nu}
        \partial_{\mu}\otimes \partial_{\nu}
    \right].
    \label{app:momentum_twist}
\end{equation}
The associated Moyal $\star$-product is
\begin{equation}
 f\star g
 =f\exp\left(
 \frac{\ii}{2}\overleftarrow\partial_\mu
 \theta^{\mu\nu}\overrightarrow\partial_\nu
 \right)g.
\end{equation}
For plane waves
\begin{equation}
    e_p(x)=e^{-ip\cdot x},
\end{equation}
one obtains
\begin{equation}
    e_p\star e_q
    =
    e^{-\frac{i}{2}p\wedge q}\,
    e_{p+q},
    \qquad
    p\wedge q
    :=
    p_{\mu}\theta^{\mu\nu}q_{\nu}.
    \label{app:plane_wave_star}
\end{equation}

Let \(G\) denote an element of the Poincar\'e Lie algebra, represented
on the algebra of fields by the map \(\rho\). Its action on a field
\(f\) is written as
\begin{equation}
    G\triangleright f
    =
    \rho(G)f.
    \label{app:representation_action}
\end{equation}

For example, translations and Lorentz transformations act on a scalar
field according to
\begin{equation}
    P_\mu\triangleright\hat{\phi}(x)
    =
    -i\partial_\mu\hat{\phi}(x),
    \label{app:translation_action}
\end{equation}
and
\begin{equation}
    M_{\mu\nu}\triangleright\hat{\phi}(x)
    =
    i\left(
        x_\mu\partial_\nu
        -
        x_\nu\partial_\mu
    \right)\hat{\phi}(x).
    \label{app:lorentz_action}
\end{equation}

The Drinfel'd twist does not modify these one-field transformation laws.
More generally, it does not modify the algebraic commutation relations
among the Poincar\'e generators or the representation carried by a
single one-particle state. Consequently, the mass-shell condition, spin
content, and one-particle dispersion relation remain the same as in the
commutative theory. In particular,
\begin{equation}
    p^2=m^2,
    \qquad
    E_{\boldsymbol{p}}
    =
    \sqrt{\boldsymbol{p}^2+m^2}.
\end{equation}

The deformation becomes relevant when a symmetry generator acts on a
product of fields or, equivalently, on a tensor product of one-particle
states. The rule for this action is encoded by the coproduct
\(\Delta\). For two elements \(G_1\) and \(G_2\) of the symmetry
algebra, the tensor-product action is defined componentwise,
\begin{equation}
\begin{split}
    &(G_1\otimes G_2)
    \triangleright
    (f\otimes h)
    \\
    &\qquad =
    (G_1\triangleright f)
    \otimes
    (G_2\triangleright h).
\end{split}
\label{app:tensor_product_action}
\end{equation}
The identity acts trivially,
\begin{equation}
    \mathbf{1}\triangleright f=f.
\end{equation}
In the ordinary theory, the action of a generator \(G\) on two field
factors is determined by the primitive coproduct
\begin{equation}
    \Delta_0(G)
    =
    G\otimes\mathbf{1}
    +
    \mathbf{1}\otimes G.
    \label{app:ordinary_coproduct}
\end{equation}
Using Eq.~\eqref{app:tensor_product_action}, one obtains
\begin{equation}
\begin{split}
    \Delta_0(G)
    \triangleright
    (f\otimes h)
    ={}&
    (G\triangleright f)\otimes h
    \\
    &+
    f\otimes(G\triangleright h).
\end{split}
\label{app:primitive_tensor_action}
\end{equation}

Since the representation space is also an algebra of $\mathcal{A}$, it is
equipped with a multiplication map
\begin{equation}
    m:\mathcal A\otimes\mathcal A\longrightarrow\mathcal A.
\end{equation}
The multiplication map and the coproduct must be compatible, meaning
that
\begin{equation}
    m\left[
        (\rho\otimes\rho)\Delta(G)(f\otimes h)
    \right]
    =
    \rho(G)m(f\otimes h).
    \label{app:coproduct_compatibility}
\end{equation}

The compatibility condition can be represented diagrammatically as
\begin{equation}
\begin{array}{ccc}
f\otimes h
& \xrightarrow{\;(\rho\otimes\rho)\Delta(G)\;} &
(\rho\otimes\rho)\Delta(G)(f\otimes h)
\\[4pt]
m\,\Big\downarrow
& &
\Big\downarrow\,m
\\[4pt]
m(f\otimes h)
& \xrightarrow{\;\rho(G)\;} &
\rho(G)m(f\otimes h).
\end{array}
\end{equation}

The primitive coproduct \(\Delta_0\) is compatible with the pointwise
multiplication map \(m_0\). It is not, however, compatible with the
twisted multiplication map
\begin{equation}
    m_\theta
    =
    m_0\circ\mathcal F_\theta^{-1}.
\end{equation}
Here and below, the action of \(\mathcal F_\theta\) on a tensor product
of fields is understood through the representation
\(\rho\otimes\rho\).

Compatibility with \(m_\theta\) is restored by deforming the coproduct
of every generator \(G\) according to
\begin{equation}
    \Delta_\theta(G)
    =
    \mathcal F_\theta
    \Delta_0(G)
    \mathcal F_\theta^{-1}.
    \label{twist_coproduct}
\end{equation}
Indeed,
\begin{align}
& m_\theta\left[
    (\rho\otimes\rho)\Delta_\theta(G)(f\otimes h)
\right]
\nonumber\\
&\quad =
m_0\left[
    (\rho\otimes\rho)
    \left(
        \mathcal F_\theta^{-1}
        \mathcal F_\theta
        \Delta_0(G)
        \mathcal F_\theta^{-1}
    \right)
    (f\otimes h)
\right]
\nonumber\\
&\quad =
m_0\left[
    (\rho\otimes\rho)
    \left(
        \Delta_0(G)\mathcal F_\theta^{-1}
    \right)
    (f\otimes h)
\right]
\nonumber\\
&\quad =
\rho(G)\,
m_0\left[
    (\rho\otimes\rho)
    \mathcal F_\theta^{-1}(f\otimes h)
\right]
\nonumber\\
&\quad =
\rho(G)(f\star h).
\end{align}

Thus, the action of \(G\) on each individual field factor remains the
ordinary one. What changes is the rule for combining these actions when
the generator acts on a product of fields or on a multiparticle state.
The twist therefore deforms the coproduct rather than the individual
one-particle representations.

The algebraic commutation relations of the Poincar\'e generators are not
changed. What is deformed is their coproduct, and hence their action on
multiparticle states. Since the translation generators commute with the
twist, their coproduct remains primitive,
\begin{equation}
    \Delta_{\theta}(P_{\mu})
    =
    P_{\mu}\otimes\mathbf{1}
    +
    \mathbf{1}\otimes P_{\mu}.
    \label{app:translation_coproduct}
\end{equation}
The Lorentz coproduct is deformed. Using
\begin{equation}
    [M_{\mu\nu},P_{\rho}]
    =
    i\left(
        \eta_{\nu\rho}P_{\mu}
        -
        \eta_{\mu\rho}P_{\nu}
    \right),
\end{equation}
one finds
\begin{align}
&\Delta_{\theta}(M_{\mu\nu})
    =
    M_{\mu\nu}\otimes\mathbf{1}
    +
    \mathbf{1}\otimes M_{\mu\nu}
    \nonumber\\
    &-
    \frac{1}{2}\theta^{\alpha\beta}
    \bigg[
        \left(
            \eta_{\mu\alpha}P_{\nu}
            -
            \eta_{\nu\alpha}P_{\mu}
        \right)
        \otimes P_{\beta}
    \nonumber\\
    &\hspace{1.5cm}
        +
        P_{\alpha}\otimes
        \left(
            \eta_{\mu\beta}P_{\nu}
            -
            \eta_{\nu\beta}P_{\mu}
        \right)
    \bigg].
    \label{app:lorentz_coproduct}
\end{align}

The theory is therefore not invariant under the ordinary Poincar\'e
action with the primitive coproduct. Rather, it is invariant under the
twisted Poincar\'e action defined by
Eq.~\eqref{twist_coproduct}. The one-particle representation and
the mass-shell relation remain unchanged, while the transformation law
of multiparticle states is deformed.

\subsection{Twisted flip and twisted statistics}
\label{app:twisted_flip}

For two identical particles in the commutative theory, particle exchange
is implemented by the ordinary flip operator
\begin{equation}
    \tau_0(\phi\otimes\psi)
    =
    \psi\otimes\phi.
    \label{app:ordinary_flip}
\end{equation}
The ordinary bosonic and fermionic projectors are
\begin{equation}
    \Pi_{\pm}
    =
    \frac{1}{2}
    \left(
        \mathbf{1}\pm\tau_0
    \right).
\end{equation}
Because the primitive coproduct is co-commutative,
\begin{equation}
    [\tau_0,\Delta_0(g)]=0,
\end{equation}
ordinary symmetrization and antisymmetrization are preserved under
Poincar\'e transformations.

This is no longer true after twisting the coproduct. In general,
\begin{equation}
    [\tau_0,\Delta_{\theta}(g)]
    \neq 0.
    \label{app:ordinary_flip_failure}
\end{equation}
Consequently, an ordinarily symmetrized or antisymmetrized state is not
mapped into a state with the same exchange symmetry by the twisted
Poincar\'e action.

The exchange operation compatible with the twisted coproduct is obtained
by deforming the flip operator with the same twist,
\begin{equation}
    \tau_{\theta}
    =
    \mathcal{F}_{\theta}\,
    \tau_0\,
    \mathcal{F}_{\theta}^{-1}.
    \label{app:twisted_flip}
\end{equation}
It obeys
\begin{equation}
    \tau_{\theta}^{\,2}
    =
    \mathbf{1},
    \qquad
    [\tau_{\theta},\Delta_{\theta}(g)]
    =
    0.
    \label{app:twisted_flip_properties}
\end{equation}
The appropriate bosonic and fermionic projectors are therefore
\begin{equation}
    \Pi_{\pm}^{\theta}
    =
    \frac{1}{2}
    \left(
        \mathbf{1}\pm\tau_{\theta}
    \right).
    \label{app:twisted_projectors}
\end{equation}
For momentum eigenstates, the twisted flip acts as
\begin{equation}
    \tau_{\theta}
    \left(
        |p\rangle\otimes|q\rangle
    \right)
    =
    e^{-ip\wedge q}
    |q\rangle\otimes|p\rangle.
    \label{app:twisted_flip_momentum}
\end{equation}
Thus exchange is accompanied by a momentum-dependent phase. For example,
a twisted bosonic two-particle state may be written as
\begin{equation}
    |p,q\rangle_{\theta,+}
    =
    \frac{1}{\sqrt{2}}
    \left(
        |p\rangle\otimes|q\rangle
        +
        e^{-ip\wedge q}
        |q\rangle\otimes|p\rangle
    \right),
    \label{app:twisted_bosonic_state}
\end{equation}
for distinct orthogonal one-particle states.

Twisted statistics is therefore not introduced as an independent
postulate. It is a direct consequence of requiring particle exchange to
be compatible with the twisted coproduct that implements Poincar\'e
covariance of the NC algebra. It is worth emphasizing that, although the twisted flip introduces a momentum-dependent exchange phase, it does not lead to genuine anyonic statistics. Indeed, $\tau_\theta^2=1$, so the exchange still furnishes a representation of the permutation group, albeit in a twisted (braided) form.

\subsection{Twisted oscillator algebra}
\label{app:twisted_oscillators}

We now pass to the second-quantized description. Let \(b_p\) and
\(b_p^{\dagger}\) denote the ordinary (bare) bosonic annihilation and creation
operators, satisfying
\begin{equation}
    [b_p,b_q^{\dagger}]
    =
    (2\pi)^3\,2E_{\boldsymbol{p}}\,
    \delta^{(3)}(\boldsymbol p-\boldsymbol q),
    \label{app:ordinary_oscillator_algebra}
\end{equation}
together with
\begin{equation}
    [b_p,b_q]=0,
    \qquad
    [b_p^{\dagger},b_q^{\dagger}]=0.
\end{equation}

The twisted (dressed) bosonic creation and annihilation operators satisfy
\begin{align}
    d_p^{\dagger}d_q^{\dagger}
    &=
    e^{ip\wedge q}
    d_q^{\dagger}d_p^{\dagger},
    \label{app:twisted_creation_algebra}
    \\
    d_p d_q
    &=
    e^{ip\wedge q}
    d_q d_p,
    \label{app:twisted_annihilation_algebra}
    \\
    d_p d_q^{\dagger}
    &=
    e^{-ip\wedge q}
    d_q^{\dagger}d_p
    +
    (2\pi)^3\,2E_{\boldsymbol p}\,
    \delta^{(3)}(\boldsymbol p-\boldsymbol q).
    \label{app:twisted_mixed_algebra}
\end{align}
The ordinary Bose algebra is recovered in the commutative limit
\(\theta^{\mu\nu}\rightarrow0\).

These relations are the second-quantized realization of the twisted
exchange operation. They show that the deformation affects the exchange
of multiparticle momentum modes, rather than the spectrum of an
individual particle.

\subsection{Dressing transformation}
\label{app:dressing_transformation}

The twisted oscillator algebra can be represented on the ordinary
bosonic Fock space through the dressing transformation
\begin{align}
    d_p
    &=
    b_p
    \exp\left(
        -\frac{i}{2}p\wedge P
    \right),
    \label{app:dressed_annihilation}
    \\
    d_p^{\dagger}
    &=
    b_p^{\dagger}
    \exp\left(
        \frac{i}{2}p\wedge P
    \right),
    \label{app:dressed_creation}
\end{align}
where
\begin{equation}
    p\wedge P
    :=
    p_{\mu}\theta^{\mu\nu}P_{\nu},
\end{equation}
and
\begin{equation}
    P_{\mu}
    =
    \int d\mu(k)\,
    k_{\mu}\,
    b_k^{\dagger}b_k
    \label{app:total_momentum}
\end{equation}
is the total momentum operator, with
\begin{equation}
    d\mu(k)
    =
    \frac{d^3\boldsymbol{k}}
    {(2\pi)^3\,2E_{\boldsymbol{k}}},
    \qquad
    E_{\boldsymbol{k}}
    =
    \sqrt{\boldsymbol{k}^2+m^2}.
\end{equation}
Using
\begin{equation}
    [P_{\mu},b_q^{\dagger}]
    =
    q_{\mu}b_q^{\dagger},
    \qquad
    [P_{\mu},b_q]
    =
    -q_{\mu}b_q,
    \label{app:momentum_oscillator_commutator}
\end{equation}
one directly obtains
Eqs.~\eqref{app:twisted_creation_algebra}--\eqref{app:twisted_mixed_algebra}.

The dressing transformation is invertible,
\begin{align}
    b_p
    &=
    d_p
    \exp\left(
        \frac{i}{2}p\wedge P
    \right),
    \\
    b_p^{\dagger}
    &=
    d_p^{\dagger}
    \exp\left(
        -\frac{i}{2}p\wedge P
    \right).
\end{align}
It is not an additional deformation unrelated to the twist; rather, it
is the Fock-space representation of the twisted coproduct and the
associated exchange structure.

We assume that the vacuum is translationally invariant,
\begin{equation}
    P_{\mu}|0\rangle=0.
    \label{app:translation_invariant_vacuum}
\end{equation}
It follows that
\begin{equation}
d_p^{\dagger}d_q^{\dagger}|0\rangle
    =
    e^{\frac{i}{2}p\wedge q}
    b_p^{\dagger}b_q^{\dagger}|0\rangle.    \label{app:two_particle_dressing}
\end{equation}
More generally,
\begin{align}
    &d_{p_1}^{\dagger}
    d_{p_2}^{\dagger}
    \cdots
    d_{p_n}^{\dagger}
    |0\rangle
    \nonumber\\
    &\qquad =
    \exp\left[
        \frac{i}{2}
        \sum_{r<s}p_r\wedge p_s
    \right]
    b_{p_1}^{\dagger}
    b_{p_2}^{\dagger}
    \cdots
    b_{p_n}^{\dagger}
    |0\rangle.
    \label{app:n_particle_dressing}
\end{align}
The twist therefore produces pair-dependent phases only in
multiparticle sectors.

The free Hamiltonian may be written as
\begin{equation}
    H_0
    =
    \int d\mu(p)\,
    E_{\mathbf p}\,
    d_p^{\dagger}d_p.
    \label{app:twisted_free_hamiltonian}
\end{equation}
Since \(p\wedge p=0\), the dressing factors cancel in the number
operator, $d_p^{\dagger}d_p
    =
    b_p^{\dagger}b_p$. Consequently, the free
Hamiltonian remains unchanged.

\subsection{The dressed scalar field}
\label{app:dressed_scalar_field}

The free real scalar field with twisted statistics is expanded as
\begin{equation}
    \hat{\phi}_{\theta}(x)
    =
    \int d\mu(p)
    \left[
        d_p e^{-ip\cdot x}
        +
        d_p^{\dagger}e^{ip\cdot x}
    \right].
    \label{app:twisted_field_expansion}
\end{equation}
The corresponding ordinary free scalar field is
\begin{equation}
    \hat{\phi}_0(x)
    =
    \int d\mu(p)
    \left[
        b_p e^{-ip\cdot x}
        +
        b_p^{\dagger}e^{ip\cdot x}
    \right].
    \label{app:ordinary_field_expansion}
\end{equation}
Substituting the dressing transformation into
Eq.~\eqref{app:twisted_field_expansion} gives
\begin{equation}
    \hat{\phi}_{\theta}(x)
    =
    \hat{\phi}_0(x)
    \exp\left[
        \frac{1}{2}
        \overleftarrow{\partial}_{\mu}
        \theta^{\mu\nu}P_{\nu}
    \right].
    \label{app:dressed_field}
\end{equation}
The derivative in this expression acts only on the field
\(\hat{\phi}_0(x)\), while \(P_{\nu}\) acts on the Fock space.
Equation~\eqref{app:dressed_field} shows that the deformation does not
modify the one-particle plane-wave solutions. Instead, the field acquires
an operator-valued dressing determined by the total momentum carried by
the multiparticle state on which it acts.

\subsection{Relation to the Bell protocol}
\label{app:twist_bell_relation}

The formal relation directly relevant for the Bell protocol is
Eq.~\eqref{app:two_particle_dressing}. If a source coherently prepares
several momentum-pair configurations, the two-particle state has the
form
\begin{equation}
    |\Psi_{\theta}\rangle
    \propto
    \sum_{i,j}
    \mathcal{A}_{ij}\,
    d_{p_i}^{\dagger}
    d_{q_j}^{\dagger}
    |0\rangle.
    \label{app:bell_twisted_state}
\end{equation}
In terms of the ordinary creation operators,
\begin{equation}
    |\Psi_{\theta}\rangle
    \propto
    \sum_{i,j}
    \mathcal{A}_{ij}\,
    e^{\frac{i}{2}p_i\wedge q_j}
    b_{p_i}^{\dagger}
    b_{q_j}^{\dagger}
    |0\rangle.
    \label{app:bell_dressed_state}
\end{equation}
For a single fixed momentum pair, the factor
\(e^{\frac{i}{2}p_i\wedge q_j}\) is a global phase and has no observable
consequence. When several momentum pairs are prepared coherently, the
same factors become relative phases between different branches of the
state.

For two momentum modes on Alice's side and two on Bob's side, the
combination invariant under independent rephasings of the local mode
bases is
\begin{align}
    \chi_{\theta}
    &=
    \frac{1}{2}
    \left(
        p_0\wedge q_0
        +
        p_1\wedge q_1
        -
        p_0\wedge q_1
        -
        p_1\wedge q_0
    \right)
    \nonumber\\
    &=
    \frac{1}{2}
    (p_0-p_1)_{\mu}
    \theta^{\mu\nu}
    (q_0-q_1)_{\nu}.
    \label{app:invariant_twist_phase}
\end{align}
This phase cannot be removed by local changes of the one-particle bases
and therefore acts as a nonlocal controlled phase between the two
momentum-mode qubits.

\section{CHSH calculation}
\label{app:chsh}

The state prepared by the source-field coupling is locally equivalent to
\begin{equation}
\ket{\Psi_\theta}
    =
    \frac{1}{2}
    \left(
        \ket{00}
        +
        \ket{01}
        +
        \ket{10}
        +
        e^{\ii\chi_\theta}\ket{11}
    \right).
    \label{eq:controlled_phase_state_CHSH}
\end{equation}
Any pure two-qubit state can be brought to Schmidt form by local unitary
transformations. Therefore, there exist local unitaries \(U_A\) and
\(U_B\) such that
\begin{equation}
    \ket{\Psi_{\mathrm S}}
    =
    (U_A\otimes U_B)\ket{\Psi_\theta}
    =
    \cos\gamma\,\ket{00}
    +
    \sin\gamma\,\ket{11},
    \label{eq:Schmidt_state}
\end{equation}
where $0\leq\gamma\leq\frac{\pi}{4}$. 
Since local unitaries do not change entanglement, the Schmidt parameter
is fixed by the concurrence,
\begin{equation}
    \sin(2\gamma)
    =
    C_\theta
    =
    \left|
        \sin\left(
            \frac{\chi_\theta}{2}
        \right)
    \right|.
    \label{eq:Schmidt_concurrence}
\end{equation}
For completeness, an explicit Schmidt-basis construction can be given
on the principal branch \(0\leq\chi_\theta\leq\pi\). Define the local
basis states
\begin{align}
    \ket{+_\theta}
    &=
    \frac{1}{\sqrt2}
    \left(
        e^{-\ii\chi_\theta/2}\ket0+\ket1
    \right),
    \\
    \ket{-_\theta}
    &=
    \frac{1}{\sqrt2}
    \left(
        e^{-\ii\chi_\theta/2}\ket0-\ket1
    \right).
\end{align}
Up to an irrelevant global phase, Eq.~\eqref{eq:controlled_phase_state_CHSH}
then becomes
\begin{equation}
    \ket{\Psi_\theta}
    \sim
    \cos\left(\frac{\chi_\theta}{4}\right)
    \ket{+_\theta}_A\ket{+_\theta}_B
    +
    \ii
    \sin\left(\frac{\chi_\theta}{4}\right)
    \ket{-_\theta}_A\ket{-_\theta}_B.
\end{equation}
The remaining factor of $i$ multiplying the second term can be
removed by a local phase shift on either Alice's or Bob's
\(\ket{-_\theta}\) mode. This gives the Schmidt form
\eqref{eq:Schmidt_state}. Thus, the Schmidt transformation requires only
local phase shifts and local two-mode mixing.

We denote the Pauli operators in the Schmidt basis by
\(\widetilde Z_A,\widetilde X_A\) and
\(\widetilde Z_B,\widetilde X_B\). For the state
\eqref{eq:Schmidt_state},
\begin{equation}
    \left\langle
        \widetilde Z_A\widetilde Z_B
    \right\rangle
    =
    1.
    \label{eq:ZZ_Schmidt}
\end{equation}
Indeed, both components \(\ket{00}\) and \(\ket{11}\) have equal local
\(Z\)-outcomes, so their product is always \(+1\).
Similarly,
\begin{align}
    \left\langle
        \widetilde X_A\widetilde X_B
    \right\rangle
    &=
    2\cos\gamma\sin\gamma=
    \sin(2\gamma)
    =
    C_\theta.
    \label{eq:XX_Schmidt}
\end{align}
The mixed correlations vanish,
\begin{equation}
    \left\langle
        \widetilde Z_A\widetilde X_B
    \right\rangle
    =
    \left\langle
        \widetilde X_A\widetilde Z_B
    \right\rangle
    =
    0.
    \label{eq:mixed_Schmidt_correlations}
\end{equation}
Alice chooses the two Schmidt-basis observables
\begin{equation}
    \widetilde A_0
    =
    \widetilde Z_A,
    \qquad
    \widetilde A_1
    =
    \widetilde X_A.
    \label{eq:Alice_optimal_settings}
\end{equation}
Bob's two observables are chosen symmetrically in the
\(\widetilde X_B\widetilde Z_B\)-plane around the
\(\widetilde Z_B\) direction:
\begin{align}
    \widetilde B_0
    &=
    \cos\varphi\,\widetilde Z_B
    +
    \sin\varphi\,\widetilde X_B,
    \\
    \widetilde B_1
    &=
    \cos\varphi\,\widetilde Z_B
    -
    \sin\varphi\,\widetilde X_B.
    \label{eq:Bob_symmetric_settings}
\end{align}
Thus, \(\widetilde B_0\) is rotated by \(+\varphi\) from
\(\widetilde Z_B\), while \(\widetilde B_1\) is rotated by
\(-\varphi\). The angle is chosen according to
\begin{equation}
    \tan\varphi
    =
    C_\theta,
    \label{eq:optimal_Bob_angle}
\end{equation}
so that
\begin{equation}
    \cos\varphi
    =
    \frac{1}{\sqrt{1+C_\theta^2}},
    \qquad
    \sin\varphi
    =
    \frac{C_\theta}{\sqrt{1+C_\theta^2}}.
\end{equation}
Equivalently,
\begin{align}
    \widetilde B_0
    &=
    \frac{
        \widetilde Z_B
        +
        C_\theta\widetilde X_B
    }{
        \sqrt{1+C_\theta^2}
    },
    \\
    \widetilde B_1
    &=
    \frac{
        \widetilde Z_B
        -
        C_\theta\widetilde X_B
    }{
        \sqrt{1+C_\theta^2}
    }.
    \label{eq:optimal_settings}
\end{align}

The observables actually implemented in the original calibrated
wave-packet basis are obtained by transforming these Schmidt-basis
operators back:
\begin{align}
    A_a
    &=
    U_A^\dagger
    \widetilde A_a
    U_A,
    \\
    B_b
    &=
    U_B^\dagger
    \widetilde B_b
    U_B.
    \label{eq:original_basis_optimal_observables}
\end{align}
Equivalently, Alice and Bob may first apply the local Schmidt-basis
rotations \(U_A\) and \(U_B\) and then perform the observables
\(\widetilde A_a\) and \(\widetilde B_b\). 

Using Eqs.~\eqref{eq:ZZ_Schmidt}--\eqref{eq:mixed_Schmidt_correlations},
the four correlators are
\begin{align}
    E(0,0)
    &=
    \left\langle
        \widetilde A_0\widetilde B_0
    \right\rangle
    =
    \cos\varphi,
    \\
    E(0,1)
    &=
    \left\langle
        \widetilde A_0\widetilde B_1
    \right\rangle
    =
    \cos\varphi,
    \\
    E(1,0)
    &=
    \left\langle
        \widetilde A_1\widetilde B_0
    \right\rangle
    =
    C_\theta\sin\varphi,
    \\
    E(1,1)
    &=
    \left\langle
        \widetilde A_1\widetilde B_1
    \right\rangle
    =
    -C_\theta\sin\varphi.
    \label{eq:optimal_correlators}
\end{align}
For the CHSH combination
\begin{equation}
    S_\theta
    =
    E(0,0)
    +
    E(0,1)
    +
    E(1,0)
    -
    E(1,1),
    \label{eq:CHSH_definition}
\end{equation}
we therefore obtain
\begin{align}
    S_\theta
    &=
    2\cos\varphi
    +
    2C_\theta\sin\varphi.
    \label{eq:CHSH_phi}
\end{align}
The choice \(\tan\varphi=C_\theta\) maximizes this expression, giving
\begin{align}
    S_\theta^{\max}
    &=
    2\sqrt{1+C_\theta^2}
    \nonumber\\
    &=
    2\sqrt{
        1+
        \sin^2\left(
            \frac{\chi_\theta}{2}
        \right)
    }.
    \label{eq:Smax}
\end{align}
This agrees with the general maximal-CHSH result for pure two-qubit
states \cite{Horodecki1995}.

The settings in Eq.~\eqref{eq:optimal_settings} are the settings that
maximize the CHSH value for a known value of \(C_\theta\). Operationally,
one may either calibrate the angle \(\varphi\) using the predicted value
of the deformation or scan over Bob's local mixing angle and determine
the maximal observed CHSH value.

\bibliographystyle{apsrev4-2}
\bibliography{ref}

\end{document}